\documentclass[letter]{aa}
\usepackage{natbib}
\usepackage{graphicx}
\usepackage{txfonts}
\def\arcsec{\hbox{$^{\prime\prime}$}}

\def\lya{Ly$-\alpha$~}
\def\lyb{Ly$-\beta$~}

\begin{document}
   \title{The SUMER \lya line profile in quiescent prominences}
   \author{W. Curdt\inst{1}
          \and H. Tian\inst{1, 2}
          \and L. Teriaca\inst{1}
          \and U. Sch\"uhle\inst{1}
          }
   \institute{Max-Planck-Institut f\"ur Sonnensystemforschung,
   Max-Planck-Str. 2, 37191 Katlenburg-Lindau, Germany\\
   \email{curdt@mps.mpg.de}
\and School of Earth and Space Sciences, Peking University, China
             }
   \date{Received December 15, 2009; accepted February 5, 2010}
\abstract
   {}
   {Out of a novel observing technique, we publish for the first time, {\it SoHO}-SUMER observations
   of the true spectral line profile of hydrogen Lyman-$\alpha$
   in quiescent prominences. With SoHO not being in Earth orbit, our high-quality data set is
   free from geocoronal absorption. We study the line profile and compare it
   with earlier observations of the higher Lyman lines and recent model predictions. }
   {We applied the reduced-aperture observing mode to two prominence
   targets and started a statistical analysis of the line profiles in both data sets.
   In particular, we investigated the shape of the profile, the radiance distribution
   and the line shape--to--radiance interrelation.
   We also compare \lya data to co-temporal $\lambda$\,1206~Si\,{\sc iii} data.}
   {We find that the average profile of \lya has a blue-peak dominance
   and is more reversed, if the line-of-sight is perpendicular to the
   field lines. The contrast of \lya prominence emission rasters is very low
   and the radiance distribution differs from the log-normal distribution of the disk. Features seen in the Si\,{\sc iii}
   line are not always co-spatial with \lya emission.}
   {Our empirical results support recent multi-thread models, which predict
   that asymmetries and depths of the self-reversal depend on the orientation of
   the prominence axis relative to the line-of-sight.}

   \keywords{Sun: UV radiation --
             Sun: filaments, prominences --
             Line: formation --
             Line: profile --
             Opacity}
   \maketitle
%
%________________________________________________________________
\section{Introduction}

Prominences protruding out of the perfect sphere of the visible solar disk are even
visible with the naked eye, when the bright disk is occulted. These enigmatic
features, which apparently withstand gravity, have attracted scientists since centuries, but despite of substantial
progress, which has been made during the last decades in understanding the
physics of prominences, important aspects are still not understood. The enormous
efforts to understand the nature of prominences is reflected in a wealth of
literature. We refer to review articles and reference material, which may be relevant
to our work \citep[e.g., ][]{Tandberg95,Spiros02,Parenti05,Wilhelm07}, and focus
on prominence observations of the hydrogen \lya line profile,
which reveal information on the physical conditions for the line formation.

Anywhere on the disk, this line is self-reversed \citep{Curdt01}.
The reversal in the profile is, among others, related to the amount of neutral
hydrogen in its ground level, which by itself is a complex function of the
temperature and density structure of the emitting plasma. In addition,
flows of the emitting or the absorbing plasma and magnetic field may modulate
the sizes of the red or the blue peak and the symmetry of the profile \citep{Curdt08,Tian09a}.

Early observations of the \lya line profile in prominences were completed with
the LPSP instrument on {\it OSO 8} \citep{Vial82} and the UVSP instrument on
{\it SMM} \citep{Fontenla88}. These photoelectric measurements had to be corrected
for the geocoronal absorption. They already have shown signatures of asymmetry
and a wide parameter range for the depth of the reversal of the profile,
features which at that time could not be reproduced by radiative transfer calculations.
Later on, the modeling work made it clear that the overall emergent profile
highly depends on the physical conditions in the prominence \citep[e.g.,][]{Gouttebroze93}.
In particular, the imprint of the incident profile and the role of the
Prominence Corona Transition Region (PCTR) were now employed to reproduce
observations \citep[e.g.,][]{Vial07}.

Recently, 2D-multithread models were established, which are based on theoretical work of
\citet{Heinzel01} and predict -- depending on the orientation of the prominence axis
relative to the line of sight (LOS) -- opposite asymmetries of the \lya and \lyb lines \citep{Gunar07,Gunar08}
and deeply or less deeply self-reversed profiles \citep{Schmieder07}.
This is the dedicated context and the rationale of our work.

\section{Observations}

Because of its wavelength range from 660~{\AA} to 1600~{\AA}, its high spectral
resolution, and its vantage point outside of the irritating geocorona,
which absorbs \lya emission, the SUMER instrument on {\it SOHO} \citep{Wilhelm95}
is ideally suited to provide information about the line profile.
Its enormous brightness exceeds however the capabilities
of the SUMER detectors and \lya can only be observed in small sections of 50 px
on both sides of the detectors beneath a 1:10 attenuating grid.
Unfortunately, the attenuation also exerts a modulation onto
the line profile, which makes it difficult to interpret such data. Attempts
made to observe \lya in quiet Sun locations on the unattenuated bare section
of the photocathode had difficulties to calibrate the local gain depression.
First results from prominence data acquired in spring 2005 have
been reported by \citet{Gunar06}, \citet{Vial07}, and \citet{Gunar08}.

In July 2008, the SUMER team found a new, unconventional method to observe the extremely
bright \lya line of neutral hydrogen with partially closed telescope aperture to
reduce the incoming photon flux. The obtained genuine \lya profiles in the quiet
Sun and coronal hole regions were analysed by \citet{Curdt08}, \citet{Tian09a}, and
\citet{Tian09b}. Here we present, for the first time, unprecedented \lya observations of
two quiescent prominences seen in June 2009 and discuss the results obtained from a
detailed analysis of the line profiles.

\begin{figure} % figure  1
%   \sidecaption
   \includegraphics[width=8cm]{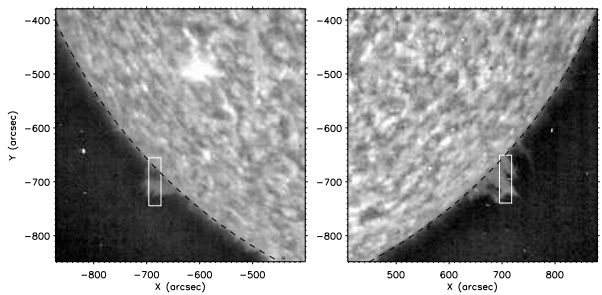}
   \caption{The prominences observed above the southeast limb on June 15 (left)
   and above the southwest limb on June 9 (right). The images are taken in the
   EIT 304 channel \citep{Boudin95} and the area of the SUMER rasters is indicated by rectangles.
   }
   \label{scanregion}
\end{figure}

\begin{figure*} % figure 2
   \sidecaption
   \includegraphics[width=12cm]{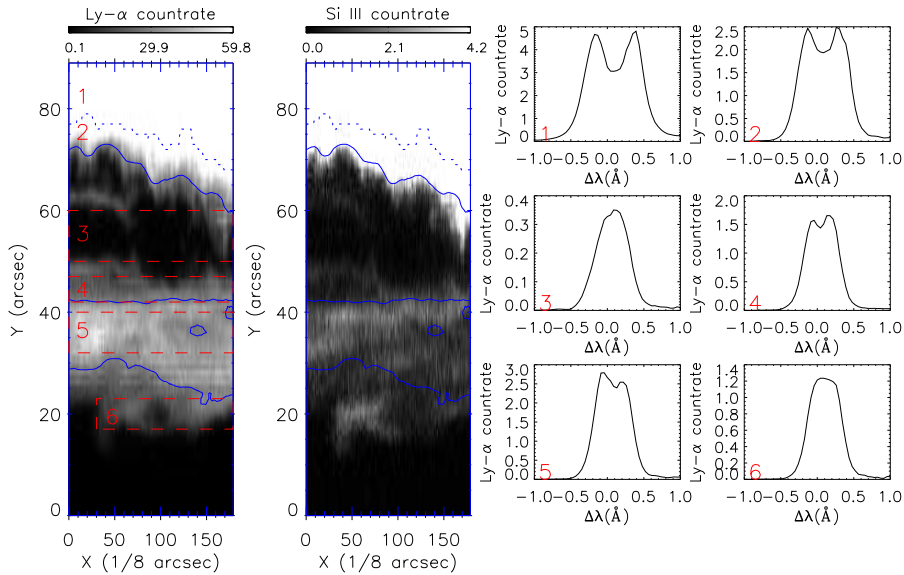}
   \caption{Raster scan in \lya and Si\,{\sc iii} of the prominence observed
   on June 15. The line radiance is given in counts/px/s.
   The raster also covers a small disk section near the southeast limb.
   Solid and dotted contours represent top 40\,\% and top 15\,\%, respectively.
   The \lya contours have been transferred to the Si\,{\sc iii} raster.
   We show the average profiles of \lya of six distinguished locations (again in instrumental units):
   \vspace{2mm}
   \newline
(1) disk,\newline
(2) limb and near disk,\newline
(3) sub-prominence void,\newline
(4) inner prominence boundary,\newline
(5) prominence core,\newline
(6) outer prominence boundary.
}
   \label{Jun15}
\end{figure*}

\begin{figure*} % figure 3
   \sidecaption
   \includegraphics[width=12cm]{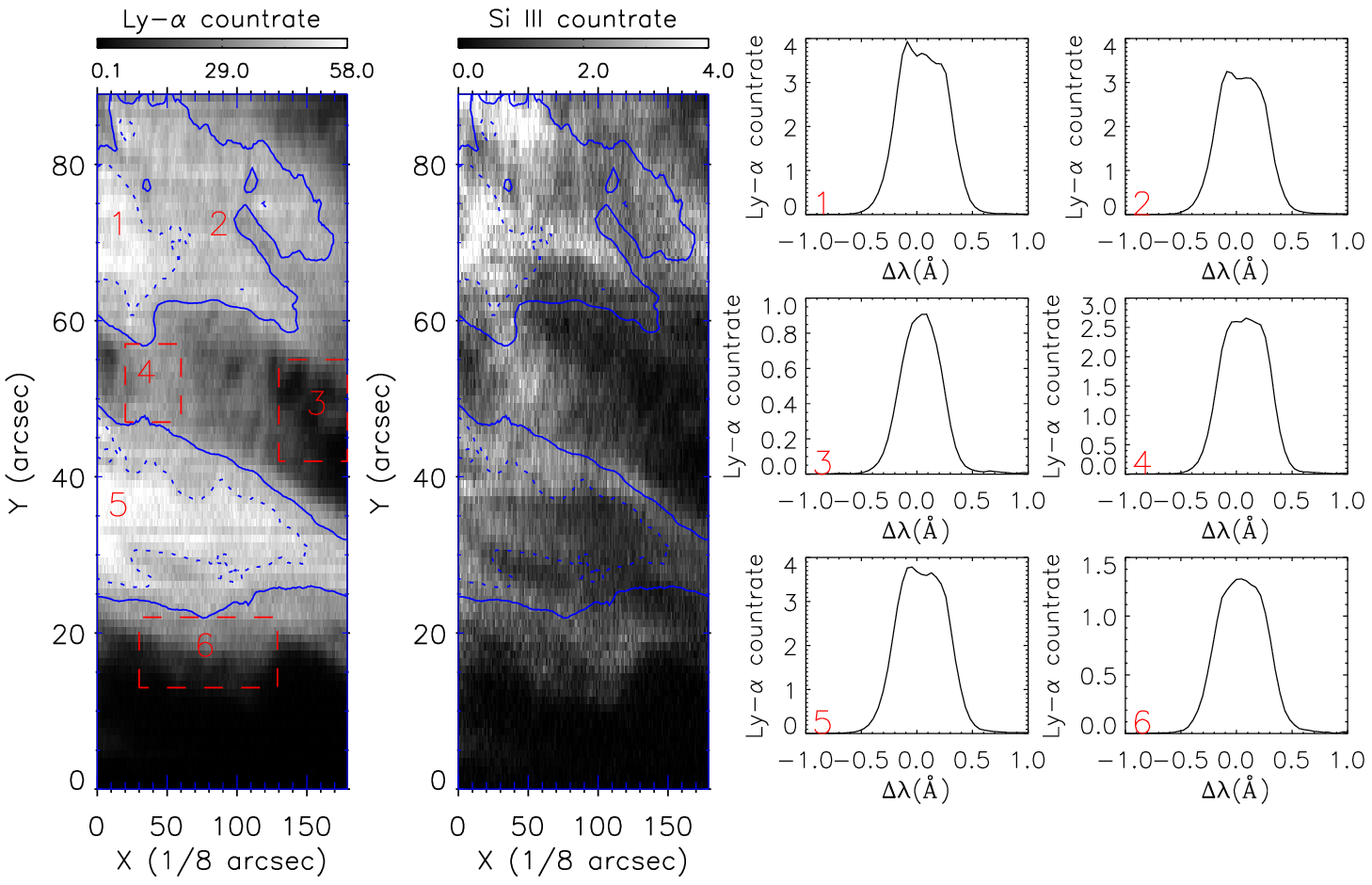}
   \caption{Idem for the observation on June 9. The raster was completed just outside the southwest limb.
   We show the average profiles of \lya of six distinguished
   locations:
   \vspace{2mm}\newline
(1) inner prominence core\newline
(2) prominence interconnection\newline
(3) sub-prominence void\newline
(4) prominence interconnection\newline
(5) outer prominence core\newline
(6) outer prominence boundary.
}
   \label{Jun9}
\end{figure*}

The new method to reduce the incoming photon flux to a moderate level,
appropriate for Ly$-\alpha$, was described in earlier work \citep{Curdt08,Tian09a,Tian09b}.
A standard procedure to partially close the door led to a reproducible
reduction to a 20\,\% level.
In June 2009, this method was applied for the first time to prominence observations.

On June 9 and 15, 2009, we completed raster scans of approximately 22\arcsec~$\times$~120{\arcsec}
at positions near the solar limb with mid-size prominences.
Two spectral windows were transmitted, 100~pixels (px) around Ly$-\alpha$ recorded on the bare
photocathode of the detector, and 50~px around $\lambda$\,1206~Si\,{\sc iii} recorded
on the KBr-coated section of the photocathode, respectively. All observations were completed with
the 0.3\arcsec~$\times$~120{\arcsec} slit. With an exposure time of 14.5~s, both lines
were observed with sufficient counts for a good line profile analysis.
For both data sets three exposures at each position were completed before the
raster was continued with a very small increment of 0.375{\arcsec}.
From a first inspection, we found no indication for temporal variations of the
object during the observing time of 45 minutes. In our statistical analysis
we keep the temporal information and assume that the sub-resolution
increment of 0.375{\arcsec} (= 3/8{\arcsec}) is equivalent to three
hypothetical increments of 1/8{\arcsec}.

The exact knowledge of the limb position and the limb distance for each pixel
is very important for prominence observations. Therefore we used additional
information, for an independent assessment of the instrument pointing uncertainty,
provided by the  hardware encoders in the instrument's housekeeping channel.
Thus, we confirmed that the azimuth movement was as expected and that the
actual east-west pointing was very close to its nominal value.
Similarly, we confirmed that in elevation the absolute positions for both
rasters differ by the nominal value of 5{\arcsec}. Since in the June 9 data set
the position of the limb can be determined, we can estimate that the overall
pointing uncertainty is in the order of 2{\arcsec} to 3{\arcsec}. The prominences as seen
in the EIT 304 channel are shown in Fig.\,\ref{scanregion},
white rectangles marking the area covered by the SUMER rasters.

Both data sets were processed with standard procedures of the
SUMER-soft library. We used the dedicated flatfield exposure of April~19, 2009
to complete the flatfield correction.

\section{Prominences in \lya and in Si\,{\sc iii}}

The rasters for both days are shown in Fig.\,\ref{Jun15} and Fig.\,\ref{Jun9}.
Note, that the x-axis also contains time information (cf., prev. section)
and that both axes are at a different scale, and that the x-dimension is stretched.
The contours delineate the top 15\,\% (dashed) and the top 40\,\% (solid) of the
pixels in the \lya brightness histogram. These contours have been transferred
to the Si\,{\sc iii} raster,

In Fig.\,\ref{Jun15} (observation on June~15) we distinguish six different segments of the raster,
separated by the blue contours or red boxes:\\
(1) disk,
(2) limb and near disk,
(3) sub-prominence void,\\
(4) inner prominence boundary,
(5) prominence core,\\
(6) outer prominence boundary.

We also display the averaged profiles of the designated areas. It is obvious
that the disk profiles are much wider and more reversed than the prominence
profiles. The disk profiles are almost symmetric, which is consistent with
the downflow argument in \citet{Curdt08} for this special geometry.
Interestingly, a clear blue-peak dominance is observed in the profile of the
prominence core.

In Fig.\,\ref{Jun9} (observation on June~9) we distinguish:\\
(1) inner prominence core
(2) prominence interconnection\\
(3) sub-prominence void\hspace{2mm}
(4) prominence interconnection\\
(5) outer prominence core
(6) outer prominence boundary.

Here, we observe just outside the limb. The contrast is even lower in this
prominence, which is shown with the same dynamic range. The profiles are
almost flat-topped, significant reversals are not seen anywhere. The blue-peak
dominance is also present in both parts of this prominence, but less evident.

The prominence is also seen in Si\,{\sc iii}. Again, the \lya radiance contours
have been transferred. These contours show, that there are considerable differences in
the  Si\,{\sc iii} spectroheliogram; structures are not co-spatial; the
prominence appears more granulated and not as diffuse as in Ly$-\alpha$. The
formation temperature of Si\,{\sc iii} is 70\,000~K, much higher than typical
prominence temperatures of 6\,000~K to 8\,000~K. Si\,{\sc iii} is a typical
transition region line. Since its wavelength is well
above 912~{\AA}, the Lyman limit, opacity effects by hydrogen can be ruled out.
The prominence is basically transparent \citep{Anzer07}. These authors also show
that the C\,{\sc i} recombination continuum below 1239~{\AA} is negligible
and, consequently, the PCTR of each unresolved thread would contribute to the
Si\,{\sc iii} emission and one would expect an appearance similar to the
cold body. The differences in appearance may be an indication for the coexistence
of hot and cold plasma with different opacities. Recent observations by {\it Hinode}-SOT
\citep{Berger08} assume buoyant bubbles of hotter plasma in quiescent prominences,
although on smaller scales. Such a scenario would also be compatible with
our observation. Without {\it Hinode}-SOT co-observations, however, our results remain
inconclusive.

\begin{figure} % figure 4
   \includegraphics[width=8cm]{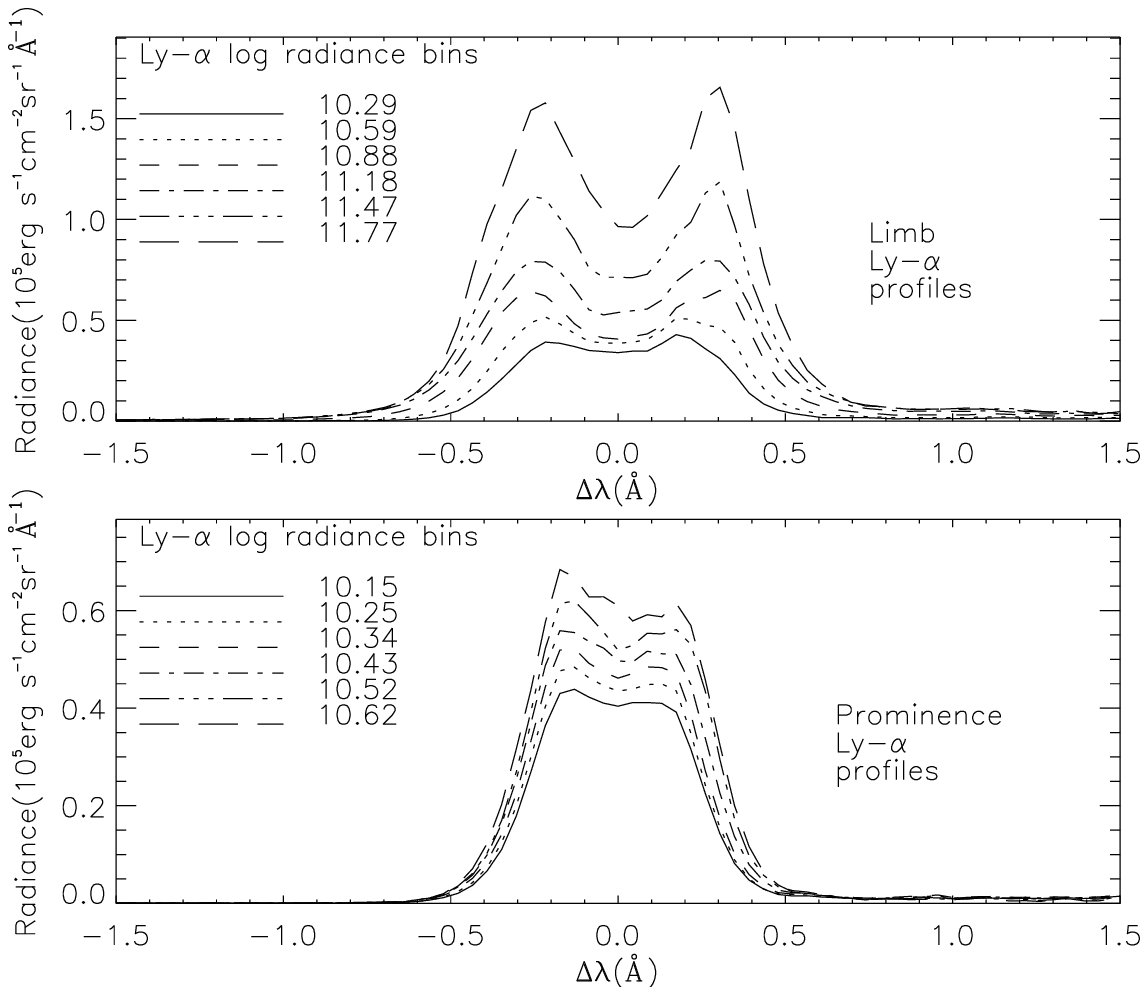}
   \caption{We have sorted the pixels within the top 40\,\% contours (solid in Fig.\,\ref{Jun15}) by their
   radiance and show the profiles of \lya of six equally spaced radiance bins.
   The northern region represents disk and limb (top), the southern region
   the prominence core (bottom).
   }
   \label{ProfJun15}
\end{figure}

\begin{figure} %figure 5
   \includegraphics[width=8cm]{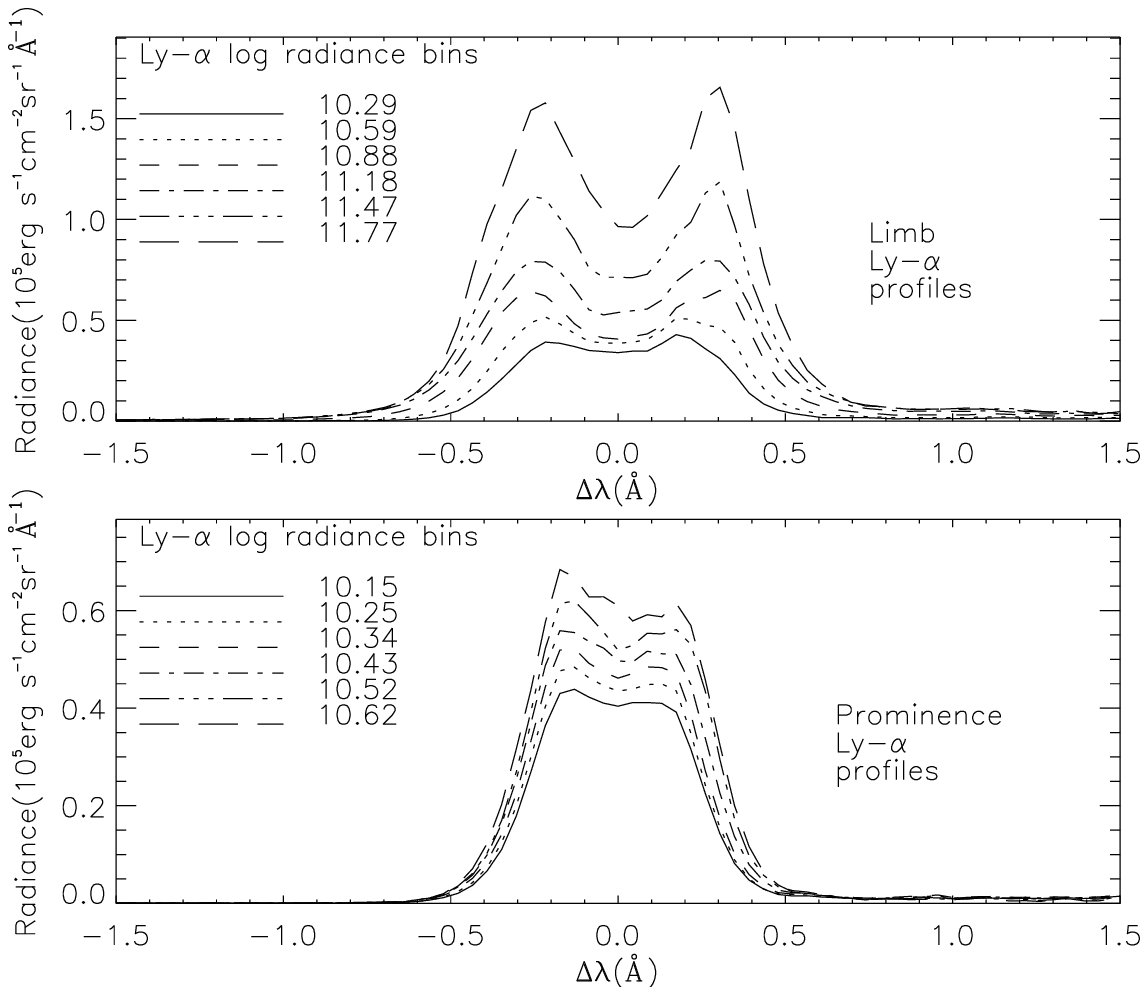}
   \caption{Idem for the inner (top) and outer (bottom) region in Fig.\,\ref{Jun9}.}
   \label{ProfJun9}
\end{figure}

We sorted the pixels of all disk locations and of all prominence locations by the
total line radiance and defined six equally spaced radiance bins.
The profiles for these bins are displayed in Figs.\,\ref{ProfJun15} and \ref{ProfJun9}.
There are striking differences of prominence profiles compared to disk profiles.
In the prominence, the contrast is much lower, reduced by a factor of 4 to 5.
The blue-peak dominance is observed in all radiance bins from the brightest
areas of the prominence core.

The central reversals of the \lya profiles in both prominences differ, the
profiles obtained on Jun 15 were more reversed than those from Jun 9. This
may be related to the different orientation of the prominence axes as derived
from EIT 304 (cf., Figs.\, 2 and 3) and Kanzelh\"ohe H-$\alpha$ images. On Jun 15,
the threads were rather perpendicular to the line-of-sight, while
more edge-on (LOS parallel to the field lines) on Jun 9. Such an
explanation would be consistent with the model calculations and predictions
of \citet{Heinzel05}. Observational evidence for such a scenario based on spectra
of the higher Lyman lines, Ly2 to Ly7, has been reported by \citet{Schmieder07}.

\section{Radiance histograms}

\begin{figure} % figure 6
   \includegraphics[width=8cm]{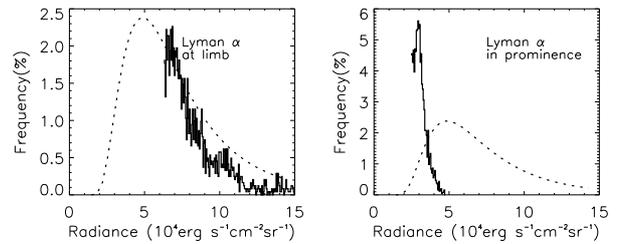}
   \caption{Radiance distribution of the limb location (1+2) in Fig.\,\ref{Jun15}
   (left) and of the prominence core (right). The low contrast of the prominence location translates
   into a narrow distribution, which differs significantly from the log-normal
   distribution, which \citet{Curdt08} found in the quiet Sun at disk center (dotted curve).}
   \label{HistJun15}
\end{figure}

We already noted the low contrast of the prominences in Ly$-\alpha$. In Fig.\,\ref{HistJun15}
we show radiance histograms of the prominence core and of the on-disk
locations in Fig.\,\ref{Jun15}. For comparison we add as dotted line the log-normal
radiance distribution of \lya in the quiet Sun as presented in earlier work \citep{Curdt08}.
Note that the disk histogram, having a different bin size, was scaled for
better comparison.
Although the small number of prominence pixels only allows a noisy distribution,
the differences are, as expected, very obvious. The histogram is per definition clipped at the
dim side because of the area selection criterion. The main difference is certainly
found in the high-radiance part; the prominence histogram completely lacks
brighter pixels, which makes it a very narrow distribution. The uniform emergent emission translates,
according to the Barbier-Eddington relation, to a uniform source function at
an optical depth, $\tau$, of unity and is indicative of homogeneous populations
of the 1s and 2p levels, and thus rather homogeneous thermodynamic conditions.

\begin{figure} % figure 7
   \includegraphics[width=8cm]{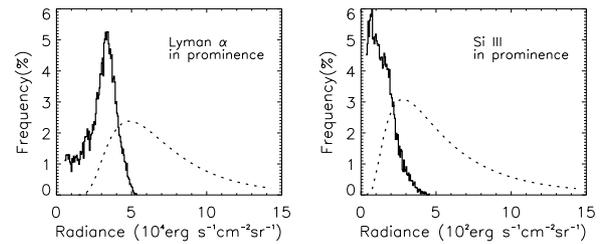}
   \caption{Idem for the prominence in Fig.\,\ref{Jun9}.
   The coronal background of pixels in the lower-15\,\% radiance category has been excluded.}
  \label{HistJun9}
\end{figure}

In Fig.\,\ref{HistJun9} we show the radiance distribution of the prominence in
Fig.\,\ref{Jun9} in \lya and in Si\,{\sc iii} emission. This data set has more
prominence pixels and allows to also include fainter pixels here.
We defined an empirically determined discrimination level to separate prominence emission
from coronal background and defined the lower-15\,\% radiance category as coronal background, which does not belong
to the prominence. The \lya histogram has a sharp upper limit and, in
contrast to the disk histogram, a low-radiance wing. The Si\,{\sc iii}
histogram of this prominence differs significantly, as one could expect, from
both the quiet Sun state (dotted line) and from its \lya counter part.

We conclude, that the radiance distributions of both prominences are, as a
consequence of dissimilar physical conditions, remarkably different from the
log-normal distribution of the average quiet Sun \citep{Fontenla88,Curdt08}.

%\newpage

\section{Summary and conclusion}

We have presented the first SUMER observations of prominences in the light of
the hydrogen \lya line at 1216~{\AA} with reduced incoming photon flux to avoid
saturation effects of the SUMER detection system. We have completed a statistical
analysis and report salient empirical results derived thereof.
As such, we found clear evidence in support of models, which predict an effect
of the orientation of the magnetic field relative to the line of sight on
the asymmetry of the \lya profile.
The Lyman lines are more reversed if the line of sight is across the prominence
axis as compared to the case when it is aligned along its axis.
Given the great variability in the appearance of prominences
and the wide range of physical parameters, the observation of two
prominences is by far not enough to cover all the issues. We felt, however, that
our results constitute a piece of information important enough to be presented
here. More joint observations of prominences and modeling of their \lya line profile are
highly desirable.

\begin{acknowledgements}
The SUMER project is financially supported by DLR, CNES, NASA, and the ESA PRODEX
Programme (Swiss contribution). SUMER is part of {\it SOHO} of ESA and NASA.
HT is supported by the International Max Planck Research School for his stay at MPS.
This non-routine observation was performed with the help of D. Germerott. This paper
greatly benefited from very constructive comments of the referee.

\end{acknowledgements}


\begin{thebibliography}{}

\bibitem[Anzer et al.(2007)]{Anzer07} Anzer,~U., Heinzel,~P., F\'arn\'ik,~F. 2007, Sol.Phys. 242, 42.
\bibitem[Delaboudini\`ere et al.(1995)]{Boudin95} Delaboudini\`ere,~J.-P.,
Artzner,~G.E., Brunaud,~J., et al., 1995, Sol.Phys. 162, 291.
\bibitem[Curdt et al.(2001)]{Curdt01} Curdt,~W., Brekke,~P., Feldman,~U., Wilhelm,~K.,
Dwivedi,~B.N., Sch\"uhle,~U., \& Lemaire,~P. 2001, A\&A 375, 591.
\bibitem[Curdt et al.(2008)]{Curdt08} Curdt,~W., Tian,~H., Teriaca,~L., Sch\"uhle,~U. \& Lemaire,~P. 2008 A\&A 492, L9.
\bibitem[Berger et al.(2008)]{Berger08} Berger,~T.E., Shine,~R.A., Slater,~G.L. et al. 2008, ApJ 676, L89.
\bibitem[Fontenla et al.(1988)]{Fontenla88} Fontenla,~J.~M, Reichmann,~E.~J, \& Tandberg-Hanssen,~E. 1988, ApJ 329, 464.
\bibitem[Gun\'ar et al.(2008)]{Gunar08} Gun\'ar~S., Heinzel,~P., Anzer,~U., \& Schmieder,~B. 2008, A\&A 490, 307.
\bibitem[Gun\'ar et al.(2007)]{Gunar07} Gun\'ar~S., Heinzel,~P., Schmieder~B., Schwartz, P.,
 \& Anzer,~U., 2007, A\&A 472, 929.
\bibitem[Gun\'ar et al.(2006)]{Gunar06} Gun\'ar~S., Teriaca,~L.,~Heinzel,~P.,~\& Sch\"uhle,~U. 2006,
in Proc.'SOHO-17. 10 Years of SOHO and Beyond', eds. H.~Lacoste and L.~Ouwehand, Giardini Naxos, Italy, ESA SP-617,63.
\bibitem[Gouttebroze et al.(1993)]{Gouttebroze93} Gouttebroze, Heinzel,~P., \& Vial,~J.-C. 1993, A\&AS, 99, 513.
\bibitem[Heinzel et al.(1987)]{Heinzel87} Heinzel et al. 1987, A\&A 183, 351.
\bibitem[Heinzel \& Anzer(2001)]{Heinzel01} Heinzel,~P. \& Anzer,~U. 2001, A\&A 375, 1082.
\bibitem[Heinzel et al.(2005)]{Heinzel05} Heinzel,~P., Anzer,~U., \& Gun\'ar~S. 2005, A\&A 442, 331.
%\bibitem[Heinzel et al.(2008)]{Heinzel08} Heinzel.,~P., Schmieder,~B., \& Farnik,~F. 2008, A\&A 686, 1383.
\bibitem[Schmieder et al.(2007)]{Schmieder07} Schmieder,~B., Gun\'ar,~S., Heinzel,~P., \& Anzer,~U., Sol.Phys. 241,53.
\bibitem[Parenti et al.(2005)]{Parenti05} Parenti,~S., Vial,~J.-C.,\& Lemaire,~P. 2005, A\&A 443, 679.
\bibitem[Patsourakos \& Vial(2002)]{Spiros02} Patsourakos,~S. \& Vial.~J.-C. 2002, Sol.Phys. 208, 253.
%\bibitem[Stellmacher \& Wiehr(2008)]{Stellmacher08} Stellmacher,~G. \& Wiehr,~E. 2008, A\&A 489, 773.
\bibitem[Tandberg-Hanssen(1995)]{Tandberg95} Tandberg-Hanssen,~E. 1995, {\it The Nature of Solar Prominences}
 Astrophys. Space Sci. Lib. 199, (Kluwer, Dordrecht).
\bibitem[Tian et al.(2009\,a)]{Tian09a} Tian,~H., Curdt,~W., Marsch, E., \& Sch\"uhle, U. 2009\,a, A\&A 504, 239.
\bibitem[Tian et al.(2009\,b)]{Tian09b} Tian,~H., Teriaca,~L., Curdt,~W., \& Vial,~J.-C. 2009\,b, ApJ 703, L152.
\bibitem[Vial(1982)]{Vial82} Vial,~J.-C., 1982, ApJ, 253,330.
\bibitem[Vial et al.(2007)]{Vial07} Vial,~J.-C., Ebadi,~H., \& Ajabshirizadeh, A. 2007, Sol.Phys. 246, 327.
\bibitem[Wilhelm et al.(1995)]{Wilhelm95} Wilhelm,~K., Curdt,~W., Marsch,~E., et al., 1995, Sol. Phys. 162, 189.
\bibitem[Wilhelm et al.(2007)]{Wilhelm07} Wilhelm,~K., Marsch,~E., Dwivedi,~B.N., \& Feldman, U. 2007, SSRv 133, 103.
\end{thebibliography}
\end{document}